\documentclass[preprint]{aastex631}
\long\def\comment#1{}
\def\kms{km~s$^{-1}$}
\def\al{Alfv\'{e}n}

\def\hinode{{\sl Hinode}}

\def\p78{{\sl P78-1}}

\def\sdo{{\sl SDO}}
\def\iris{{\sl IRIS}}

\def\psp{{\sl PSP}}
\def\so{{\sl Solar Orbiter}}

\def\fexii{Fe~{\sc xii}}

\def\heii{He~{\sc ii}}

\def\al{Alfv\'{e}n}

\def\kms{km~s$^{-1}$}

\def\etal{et~al.}

\begin{document}
%

\title{How Small-scale Jet-like Solar Events from Miniature Flux Rope Eruptions Might Produce the Solar Wind}

\author{Alphonse C. Sterling}
\affiliation{NASA/Marshall Space Flight Center, Huntsville, AL 35812, USA}

\author{Navdeep K. Panesar} 
\affiliation{Bay Area Environmental Research Institute, NASA Research Park, Moffett Field, CA 94035, USA}
\affiliation{Lockheed Martin Solar and Astrophysics Laboratory, 3251 Hanover Street, Building 252, Palo Alto, CA 94304, USA}

\author{Ronald L. Moore} 
\affiliation{Center for Space Plasma and Aeronomic Research, \\
University of Alabama in Huntsville, Huntsville, AL 35805, USA}
\affiliation{NASA/Marshall Space Flight Center, Huntsville, AL 35812, USA}

\comment{

\author{\etal}

\author{Louise K. Harra} 
\affiliation{Physikalisch Meteorologisches Observatorium Davos, World Radiation Center, 7260 Davos, Switzerland}
\affiliation{Institute for Particle Physics and Astrophysics, ETH Z{\"u}rich, 8092 Z{\"u}rich, Switzerland}

} 

\begin{abstract}

We consider small-scale jet-like events that might make the solar wind, as has been suggested in recent studies.  We show that
the events referred to as ``coronal jets'' and as ``jetlets'' both fall on a power-law distribution that also includes large-scale eruptions
and spicule-sized features; all of the jet-like events could contribute to the solar wind.  Based on imaging and magnetic field data,
it is plausible that many or most of these events might form by the same mechanism: Magnetic flux cancelation produces small-scale flux
ropes, often containing a cool-material minifilament.  This minifilament/flux rope erupts and reconnects with adjacent open coronal field,
along which ``plasma jets" flow and contribute to the solar wind.  The erupting flux ropes can contain twist that is transferred to
the open field, and these become \al ic pulses that form magnetic switchbacks, providing an intrinsic connection between 
switchbacks and the production of the solar wind.

\end{abstract}


\section{Introduction}
\label{sec-introduction}

With the advent of fresh data from satellites in the near-Sun heliosphere, there is renewed interest in the possibility
that small-scale eruptions might be the cause of the pervasive solar wind outflow.  \citet{raouafi.et23} suggest that
small-scale jet-like outflows, called {\it jetlets}, might be the primary responsible agent.  EUV images show that
jetlets have widths of a few 1000\,km, and are extremely abundant over the entire quiet- and coronal-hole Sun.  
Different from active regions, they are 
present throughout the solar cycle, as is the solar wind.  

\citet{raouafi.et23}  further speculated that a subset of
the copious small-scale magnetic cancelation episodes occurring in the photosphere is responsible for driving 
the jetlet production through magnetic reconnection.  They argue that their estimated rate of 
$5 \times 10^5$\,jetlets\,day$^{-1}$ would be sufficient to supply the mass and energy requirements of the
solar wind.  Furthermore, they argue that the magnetic switchbacks,
which are localized rotations in the solar wind magnetic field that are ubiquitously detected in near-Sun 
Parker Solar Probe ({\psp) data, could also be a consequence of the magnetic reconnection events that produce the
jetlets and solar wind, and are built up and triggered by fine-scale flux cancelation.

\citet{chitta.et23}, on the other hand, use \so\ data to conclude that there are
even smaller-scale features, which they call ``picoflare jets," with widths $\sim$100\,km.  These observations are 
of a coronal hole using \so's EUV High Resolution Imager (HRI$_{\rm EUV}$) at 174\,\AA, and were taken 
during a close approach of 0.332 AU, where the images had a spatial resolution of about 237\,km.  Based on 
their morphology, they suggest that these picoflare jets are driven by magnetic reconnection.  Based on the 
number of events that they observed, and extrapolating the filling factor of the jets over the area they observed
to the entire Sun, they estimate that the picoflare jets might account for 20\% of the solar wind mass flux.

Both of these studies present observations in support of small-scale jet-like events (which we will call ``small-scale jets") 
providing mass and energy to the solar wind.  In this work, we present a picture for how the small-scale jets come from 
built-up magnetic energy, 
are generated, produce outflows, and can propagate into the solar wind and often form switchbacks.  We also
speculate on how these small-scale jets could contribute mass and heating to the solar wind.

A previous study, \citet{moore.et11}, also suggested that small-scale jets could lead to generation of the 
solar wind and the entire heliosphere, but where they took those jets to be type~II spicules.  This work 
updates the concept of that idea, based on much new information that we have gained on jets and 
jet-like features, and we consider other subsequently discovered jet-like features than spicules (e.g., jetlets), 
although spicules might still be a contributing component.  That older \citep{moore.et11} work was based on 
previous ideas for how X-ray coronal jets 
were thought to form at the time, namely via emerging magnetic field reconnecting with surrounding ambient 
coronal field \citep{shibata.et92,yokoyama.et95}.  Much evidence now supports that coronal jets are instead produced by
small-scale filament (minifialment) eruptions, and that those minifilament eruptions are at least often a consequence of 
magnetic flux cancelation.

\section{Coronal Jets}
\label{sec-jets}

Coronal jets are usually observed at soft X-ray (SXR) and/or EUV wavelengths.  In SXR images, jets reach about 50,000\,km
with widths $\sim$8000\,km, based on observations from \hinode's X-ray telescope (XRT) by \citet{savcheva.et07}.  
Extrapolating their values of 60 jets/day in the two coronal holes yields an occurrence rate of a few hundred per day 
over the entire solar surface for the size and quality of jets that they observed in that wavelength band.  They have lifetimes
of from about ten to 30 minutes, and estimates for the energy expended in coronal jets range over 
$\sim$10$^{26}$---$10^{29}$\,erg \citep[e.g.,][]{raouafi.et16,hinode.et19,sterling.et23a}.

\citet{sterling.et15} argued that essentially all jets result from the eruption of a minifilament.  These minifilaments are frequently
seen in absorption in Solar Dynamics Observatory (\sdo) Atmospheric Imaging Assembly (AIA) EUV images, often including in the
\heii\ 304\,\AA\ channel, indicating that 
they are likely cool (chromospheric-temperature) features that can reside in the low corona, a~la typical filaments.  In the following,
we describe jet production in an open magnetic field environment, since we are focusing on jets that might contribute to the
solar wind.  This obviously applies to a coronal-hole environment, although the same arguments can be extended 
to quiet Sun and even active regions \citep{panesar.et16a,sterling.et23b}. 

 In an open-field location such as a coronal hole, the background field will largely 
be unipolar, consisting of a single majority polarity.  The minifilament forms at a location 
where a minority-polarity flux patch resides inside of the sea of surrounding majority polarity, with the resulting magnetic 
topology being that of an anemone region  \citep{shibata.et07}; that is, the minority polarity fans out into a three-dimensional 
lobe connecting to the surrounding majority-polarity field, with a magnetic null point elevated above the minority flux patch, and with
the surrounding coronal field forming a pseudostreamer magnetic configuration as the envelope of the anemone. 
Panels~(a)---(c) in Figure~\ref{s_m20_matome_zu} (below) show a sketch of the eruptive process.  In this two-dimentional sketch,
the anemone's base appears as a double lobe (in the 2D cross-section).  Prior to eruption, the minifilament sits in one of the 
lobes, along a magnetic 
neutral line between the majority and minority polarities.  Because the minifilament looks like a scaled-down version of a typical
solar filament, \citet{sterling.et15} assumed that upon eruption the cool minifilament material would be wrapped inside of an
erupting minifilament magnetic flux rope.  It is this flux rope eruption that is key to the jet formation, although the cool minifilament material
is vital to understanding the evolution of that field leading to and during the eruption. 

Upon eruption, the 
minifilament/flux rope is expelled toward the null over the minority polarity, and undergoes reconnection
with the far-side ambient coronal field.  This reconnection is of the ``interchange" variety, which was 
called {\it external reconnection} in \citet{sterling.et15}, with previously closed field of the
flux rope becoming new open field along the pseudostreamer spire and new closed field over the lobe of the anemone 
opposite to where the minifilament eruption originated; this outflowing heated material becomes the spire of the
jet that is observed in SXRs and/or EUV\@.  If the minifilament/flux rope erupts far enough, the external
reconnection can erode away enough of the enveloping flux rope field so that the cool minifilament material can also 
escape and flow outward along the spire; this results in a cool component to the jet, often seen in AIA 304\,\AA\ images
\citep{moore.et10,moore.et13}.  Eruptions of miniature filaments at the start of jets had been seen in 
previous investgations also \citep[e.g.,][]{nistico.et09,shen.et12,adams.et14}

About concurrent with \citep[or slightly before or slightly after; see][]{moore.et18} the external reconnection, {\it internal reconnection} 
also occurs among the legs of the erupting minifilament field that are still rooted in the solar surface.  This corresponds to
the flare-producing reconnection below typical erupting filaments.  In the case of the erupting minifilament, a strong 
brightening is often apparent in SXRs, occurring off to the side of the jet base from which the minifilment erupts.
This brightening, first identified by \citet{shibata.et92}, was identified as a miniature flare by \citet{sterling.et15}, who called the
feature a jet bright point, or JBP\@.

\citet{panesar.et16a} examined the cause of jet-producing minifilament eruptions, by tracking the magnetic base of ten on-disk 
quiet Sun jets that were observed in AIA images.  Using \sdo/Helioseismic and Magnetic Imager (HMI) magnetgrams,
they found that in all their cases flux cancelation occurred in the jet base before and during the jetting time.
Similarly, \citet{panesar.et18a} studied 13 on-disk coronal hole jets, and again found pre-eruption and during-eruption 
flux cancelation to occur at the jet 
locations.  These findings are consistent with results from several single-event earlier studies 
\citep[e.g.,][]{shen.et12,young.et14a,young.et14b,adams.et14}, and also with several later studies 
\citep[][]{mcglasson.et19,chen.et20,muglach21}.  Thus, there is strong evidence that cancelation leads to at least 
a substantial portion of coronal jets.


Coronal jets, then, are smaller-scale analogues to larger-scale typical solar eruptions.  While large-scale filaments erupt to make
a typical solar flare and often a coronal mass ejection (CME), minifilaments erupt to make a JBP and a jet spire.  
\citet{sterling.et18} have examined this issue from the other direction by presenting evidence that relatively magnetically isolated 
CME-producing eruptions do indeed appear to behave as larger-scale jet-producing minifilament eruptions.


Several studies indicate that jets show spinning motion as they extend outward from the surface, including \citet{patsourakos.et08,raouafi.et10,sterling.et10a,curdt.et12,morton.et12,shen.et12,chen.et12,hong.et13,joshi.et18,zhelyazkov.et18,liu.et19,panesar.et22}.
 This twisting/untwisting motion has been supported with spectroscopic studies also \citep{pike.et98,kamio.et10}.  \cite{wang.et98}
 found that some jets can persist well into the corona, manifesting in white-light coronagraphs as ``white-light jets," 
which are sometimes called ``narrow CMEs." \citep{sterling18}.  \citet{moore.et15} found that the jets from which the white-light jets
originated were ones that tended to have a larger amount of untwisting as they ascended, as measured using AIA 304\,\AA\ movies, 
and they also found evidence that the (un)twisting of the jets persisted as an oscillatory swaying movement of the white-light jets.

We can synthesize the flux cancelation and jet rotation into the minifilament-eruption picture for jets as follows: The flux cancelation
results in the formulation of a magnetic flux rope. If conditions are appropriate, then a cool-material minifilament forms along
that flux rope \citep[see][]{panesar.et17}.  If the cancelation occurs along a sheared-field neutral line, then that shear can be 
converted into twist in the resulting minifilament/flux rope \citep{vanball.et89}.  As the cancelation continues \citep[or the cancelation 
might resume after 
it had paused][]{panesar.et17},
the flux rope becomes unstable and erupts outward, leading to the jet as described above.  The twist of the flux rope can then
be transmitted to the open field through the external reconnection, following a process described by \citet{shibata.et86}.

Several reviews and summaries of jets are now available \citep{shimojo.et00,shibata.et11,raouafi.et16,hinode.et19,shen21,sterling21,schmieder22,sterling.et23a}.

\section{Jetlets}
\label{sec-jetlets}

\citet{raouafi.et14} identified features in AIA 171 and 193 movies that appeared similar to coronal jets, but of shorter durations 
-- tens of seconds to a few minutes -- and of smaller size than typical coronal jets.  That study found these jetlets to be obvious at the 
base of solar coronal plumes.  

\citet{panesar.et18b} also studied jetlets, using both AIA EUV images and UV images from the  
Interface Region Imaging Spectrograph (\iris) satellite.  They found the jetlets to appear in 
more general network regions rather than just at the base of plumes, and to have typical lengths of $\sim$27,000\,km,
widths of 3000\,km, and lifetimes of 3 minutes. Using high-resolution ($0.''129$ pixels,
compared to, e.g., $0.''6$ pixels for AIA) Hi-C2.1 EUV 172\,\AA\ images \citep{rachmeler.et19},  \citet{panesar.et19} observed 
six jetlets that were of even smaller size: about 9000\,km in length and 600\,km in width.  These Hi-C jetlets also were rooted 
at the edges of magnetic network lanes.  

\citet{raouafi.et23} estimated the rate of energy expenditure of a jetlet to be 
$\sim$5$\times 10^{22}$\,erg~s$^{-1}$.  Assuming lifetimes of 20\,s to 5 minutes \citep[cf.][]{raouafi.et14,panesar.et18b}, 
this gives a total energy of $\sim$10$^{24}$---10$^{25}$\,erg for a single jetlet.

All four of the studies: \citet{raouafi.et14}, \citeauthor{panesar.et18b}~(\citeyear{panesar.et18b},\citeyear{panesar.et19}), and \citet{raouafi.et23}, 
used magnetograms to investigate the magnetic
behavior at the base of the jetlets, and all found evidence that magnetic cancelation episodes were closely tied to jetlet formation.
For example: \citet{raouafi.et14} conclude that the jetlets result from ``flux emergence followed by magnetic cancellation of the minority polarity 
with the dominant unipolar field concentration."  From a study of ten jetlets, \citet{panesar.et18b} found clear evidence for flux 
cancelation preceding nine of them, with an average rate of about $1.5 \times 10^{18}$\,Mx\,hr$^{-1}$.   \citet{panesar.et19} found
that four of their six jetlet-like events resulted from flux cancelation.  

Furthermore, \citeauthor{panesar.et18b}~(\citeyear{panesar.et18b,panesar.et19}) argued that jetlets are analogues 
of the larger-scale coronal jets discussed in \S\ref{sec-jets}.  Jetlets are similar to coronal jets in that they appear jet-like, viz.\ with a
bright base and a spire that extends in time.  Moreover, their extension velocity in EUV is $\sim$70\,\kms, which is
similar to the corresponding coronal jet value of $\sim$100\,\kms\ and 70\,\kms\ for quiet Sun and coronal hole
jets, respectively  \citep{panesar.et18b,panesar.et16b,panesar.et18a}.  The flux cancelation at the base of jetlets
mimics that for many jets.  Also similar to jets, some jetlets might display twisting motion 
\citep{panesar.et18b,panesar.et19}, although these observations are at the limit of detection and therefore not conclusive.  

Despite searching for
erupting minifilments at the base of jetlets, however, the Panesar~\etal\ studies did not find any.  This could well be due to the
small size of the features.  Even in jets, erupting minifilaments can be difficult to detect in the smaller and less distinct 
ones, especially the so-called ``standard jets"; these are jets with relatively narrow spires, compared to the size of the 
same jet's base region, especially when observed in SXRs \citep[][]{moore.et10,moore.et13}. These standard jets 
often form from erupting minifilaments that are confined to the base of the jet \citep{sterling.et22}, and thus would be more difficult to 
detect than in jets with erupting minifilaments that are ejected out along the jet's spire (often forming ``blowout jets," which have broad
spires compared to the base in SXRs).   \citet{kumar.et22}, however, did report observing
a minifilament that apparently is erupting in a jetlet that they report is about two times larger than a typical jetlet but smaller than typical jets.  

Therefore it might be that higher-resolution EUV imaging \citep{sterling.et23a} will be required to confirm the presence of 
erupting minifilaments in most
jetlets.  With this caveat then, we regard the above-noted similarities between jetlets and coronal jets as evidence
that jetlets are smaller-scale versions of coronal jets.  


\section{Smaller Jet-like Features}
\label{sec-smaller_jets}

If jetlets are indeed small-scale jets, then jet-like features likely occur on still smaller size/energy scales also.  

Higher resolution instruments should reveal such features, if they exist.  Indeed, \so\ identified small-scale features 
dubbed ``campfires"  \citep{berghmans.et21,zhukov.et21}. These features are described as small-scale short-lived 
coronal brightenings that can appear loop-like, dot-like, or complex structures, and that live for 1---60\,min.  

\citet{panesar.et21} studied 52 random campfires, and concluded that: they are ``rooted at the the edge of photospheric 
[magnetic flux] lanes," most appear above magnetic neutral lines between opposite-polarity magnetic patches, and that most of them
are ``preceded by a cool-plasma structure, analogous to minifilaments in coronal jets."  They also conclude that some
of the campfires appear similar to coronal jets.  From their table of 52 events, they list nine of them as appearing jet-like.
Culling out the properties of these nine, we find that those jet-like campfires have lengths of $5500\pm2300$\,km.  This 
is not far different from the 
statistics on all 52 given in \citet{panesar.et21} $4500\pm2500$\,km, and so this indicates that the remaining 43 events
that are not classified as ``jet-like" may be fundamentally similar.  These lengths are somewhat smaller than 
the average values for jetlets, and therefore they could be part of a population of smaller-sized objects that operate via
the same mechanism as jets.  The dark feature visible in them appears to be an erupting minifilament, like those seen in
coronal jets.

Figure\,\ref{eui_nkp_campfire13_crop} shows images of a jet-like campfire from the \so\ Extreme Ultraviolet Imager (EUI) High Resolution Imager 
HRI$_{\rm EUV}$, at174\,\AA\@.  This is event\,13 in \citet{panesar.et21} (see Fig.\,6 of that paper).  Arrows in the figure point out the
 erupting dark feature.  This feature is also prominent in absorption in AIA 304, 171, and 193\,\AA\ images, suggesting that it
 is composed of cool material, and clearly is erupting in this sequence 
 \citep[animations are available in][]{panesar.et21}.  The look and dynamics of this absorbing feature are essentially idential
 to the minifilaments that erupt to make coronal jets, and therefore this is in all likelihood an event occurring via the same
 jet-producing mechanism, but on a size scale smaller than that of typical jets.

The \so-observed picoflare jets of \citet{chitta.et23} are much smaller still, with spatial scales of $\sim$few $\times 100$\,km, which 
is comparable to the widths of spicules.  They estimate a lower limit of the kinetic energy to be $\sim$10$^{21}$\,erg for the picoflare
jets.  Spicules might require $\sim$10$^{25}$\,erg, but that is based on estimates for the gravitational energy \citep{sterling00} and
so a direct comparison with the quoted picoflare energy value is likely not appropriate.  

\citet{sterling.et16a} and \citet{sterling.et20b} discuss in some detail the possibility that some spicules might be made
by the coronal jet mechanism \citep[also see][]{samanta.et19}.  Spicules have morphological differences with coronal jets 
(and jetlets), however. For example, 
a bright base, common in the jet-like features, is not obvious in spicules, although it is
possible that the internal reconnection responsible for that brightening in the jets does occur at the base of the spicules, but not
enough photons are able to radiate through
the dense chromosphere for a brightening to appear \citep{sterling.et20b}.  

Also, there is as yet no convincing observation of an erupting minifilament
at the base of spicules (although \citeauthor{sterling.et20b}~\citeyear{sterling.et20b} point out candidate detections).  Instead, 
most spicules seem to form very low down, and appear as a spicular outflow at earliest detection.  Some active region jets
also have such an appearance, like geysers \citep{paraschiv.et19,paraschiv.et20,paraschiv.et22}.  But nonetheless, in the
active region jet case there is evidence that they do indeed result from minifilament eruptions, but where the external 
reconnection occurs at a low altitude and
is often obscured by surrounding elevated low-atmosphere material \citep{sterling.et23b}.  Similarly, an erupting putative 
{\it micro}filament might make spicules but be hidden by surrounding chromospheric material.  Future observations, perhaps assisted
by numerical simulations, will be required to determine whether the coronal-jet mechanism makes some or most spicules, or if they are
created instead by one or more different mechanisms \citep[e.g.,][]{martinez-sykora.et17,iijima.et17,kotani.et20}.


\section{Size Distributions of Erupting Filament-like Features, Revisited}
\label{sec-distribution}

Because jets appear to be caused by smaller-scale versions of filament eruptions that make typical solar flares and CMEs,
\citet{sterling.et16a} considered whether the jet-production mechanism might also occur on smaller size scales, with a power-law-type
distribution.  They specifically addressed this in terms of whether the same mechanism might make some or most solar spicules.
They plotted the size of the erupting filament feature on the abscissa; the largest of these are the filament eruptions that make
flares and CMEs, using available values for the sizes of filaments.  The next smallest feature they plotted was for minifilaments
that erupt to make jets, based on the sizes provided by \citet{savcheva.et07}.  For the size of the erupting microfilaments that 
they postulated might make spicules, they used measured values of spicule widths; this is because the widths of jet spires for
polar-coronal-hole \hinode/XRT-observed jets are roughly in agreement with the size of the erupting minifilaments that made 
a different set of polar-coronal-hole \hinode/XRT-observed jets used in \citet{sterling.et15}.  For the ordinate, they plotted the
estimated number of the events occurring on the entire Sun at any given time.  Such estimates for spicules are available from
historical studies.  For jets, they estimated values using rates in polar coronal holes given by \citet{savcheva.et07}.  
Similarly, flare and CME rates, along with flare durations, were used to make estimates for the number of large-scale eruptions
occurring on the Sun at any time \citep{veronig.et02,yashiro.et04,chen11}.  Considering the extent of the ranges of the
values for the measured or estimated quantities (which take the place of ``error bars" on the plot), a best-fit line to all three of 
these points showed that those three values are consistent with following a power law.  This shows that the idea that
some percentage of spicules being made with the coronal-jet mechanism is consistent with a power-law scaling of
eruptions: eruptions of smaller filament-like features become more numerous as the size of the erupting feature gets
smaller.

This idea was really based on an extension to smaller size scales of the filament-eruption mechanism making two types
of features: typical flare- and CME-producing filament eruptions, and minifilament eruptions that make coronal jets.  This
is because, as mentioned earlier, spicules have some morphological differences from the larger eruptions, and also because
erupting microfilaments have yet to be convincingly observed.  There should, though, be other jet-like features between 
the coronal jets and spicules that could be added to the plot, since there is considerable size difference between the minifilaments
that erupt to make jets ($\sim$8000\,km), and the expected size of potential microfilaments that might erupt to make some spicules
(a few 100\,km).  Jetlets fall into that intermediate size range.  Until now there have not been reliable counts of the possible number
of jetlets on the Sun, but recently such an estimate has become available, and so we can add a new point to our size-distribution 
plot.

We require the size of erupting minifilaments that might make jetlets.  As discussed above (\S\ref{sec-jetlets}), there has been only 
one reported observation of an erupting minifilament making a jetlet \citep{kumar.et22}, but there have been observations of 
dark absorbing features in the jet-like campfires \citep{panesar.et21}, and those campfires are only slightly smaller than jetlets.  Moreover, 
jetlets have morphological 
similarities to coronal jets, and also have magnetic-field behavior at their base that is similar to that of coronal jets (i.e.,
frequently showing cancelation).  Therefore, it is plausible to speculate that many jetlets result from small minifilament eruptions.  
Again taking the size of those expected erupting minifilaments to be about the width of a jetlet spire, we merely have to consider
the observed width of those spires.  From \citet{panesar.et18b}, the quoted range is $3200\pm 2000$\,km.  From 
\citep{kumar.et22}, the quoted values are $\sim$2$''$---3$''$.  \citet{panesar.et19} observed smaller-sized jetlets
with Hi-C's higher resolution, and found spire widths of $600\pm 150$\,km.  Based on this, we adopt a range
of values for the potential erupting minifilaments that might make jetlets to be over the range 500---5000\,km.
 
For the number of jetlets on the Sun, we rely on the values given in \citet{raouafi.et23}.  They looked at a fixed field of view (FOV)
of $70'' \times 70''$ magnetogram observations from Big Bear Solar Observatory's Goode Solar Telescope (BBSO/GST), and counted the
number of cancelation events that they saw over that FOV (88) over an approximately 90-minute-long nearly continuous observation 
period. During this time, they detected three EUV jetlets in the FOV\@.  Extrapolating this to the entire Sun, \citet{raouafi.et23} 
concluded that there are six jetlets~s$^{-1}$ initiated over the entire Sun.  If as above (\S\ref{sec-jetlets}) we take the lifetime
of jetlets to range between 20---300\,s, then this yields a total number of jetlets on the Sun at any instant to be 120---1800.

Figure~\ref{fourpt_dist_zu} shows the resulting four-point plot, where the horizontal and vertical lines around the point represent
the full extent of the above-mentioned ranges for the erupting-minifilament sizse and the total number of events on the Sun
at any instant.  The plotted solid line is the same fit as in \citet{sterling.et16a}, that is, it is a best-fit line to the three points
excluding the jetlets (i.e., the first, third, and fourth point, measured from left to right on the abscissa).  Considering the
range bars, the  determined value for jetlets fits on this line.  Therefore, the jetlets are consistent with fitting on the
distribution of eruptive filament-like events, spanning large-scale eruptions that make typical solar flares and CMEs,
down to some percentage of the spicules.  

Recently, \citet{uritsky.et23} have published results of a quantitative investigation of the occurrence rate of coronal outflows 
involving 2300 events. They look at the size scale of the features leaving the Sun above a polar coronal hole in AIA 171\,\AA\
images over a six-hour period.  They find size scales (``transverse size") of the outflows to range from the smallest detectable sizes up to 
$\sim$4$ \times 10^4$\,km, and they show that the bulk of the outflows follow an approximate power-law distribution in their 
occurrence rate versus the sizes of the outflows.  Our plot has a different vertical axis from theirs (we plot the number 
of erupting events on the Sun at a given time, while they plot the occurrence rate), and so we cannot make a direct comparison
of the distribution of events that they see with our results in Figure\,\ref{fourpt_dist_zu}.  

We can, however, crudely compare the \citet{uritsky.et23}
event numbers with ours, by estimating the instantaneous number of their events that would be present over the entire Sun based on what they
observe in their limited field of view. 
From Fig.\,5(b) of \citet{uritsky.et23}, most of their 2300 events have a transverse size of between about 500\,km and 3000\,km,
based on the half-width of the transverse-size distribution plotted in that figure; we can use these values as range-bar lower and
upper limits, and take the value to be 1750\,km for our Figure\,\ref{fourpt_dist_zu} abscissa.  For our ordinate, we have to 
estimate/guesstimate the number of \citet{uritsky.et23} events that would occur over the entire Sun at a given time.  They
observed 2300 events in six hours, from a portion of the solar limb that extends for $1/8$-th of the solar 
circumference.  A substantial unknown, however, is from how far along the Earth-Sun line-of-sight the events observed above
the limb by \citet{uritsky.et23} originate.  That is, among the 2300 events observed in the plane-of-sky at the limb, some will originate
from exactly at the limb, while others will originate from somewhat inside the limb, and others will originate from somewhat beyond the limb;
we do not know from how far inside and how far beyond the limb those features might originate and be observed in the observations of 
Fig.\,1(a) of \citet{uritsky.et23}. To obtain a rough minimum estimate of the whole-Sun instantaneous number of events, we 
will assume that the maximum for this line-of-sight-contribution 
region's extent is $1/8$-th of the circumference on either side of the limb.  We can then approximate this source region of the \citet{uritsky.et23} features to be a two-dimensional
rectangle with one side $C_{\odot}/8$ in length (this is the along-the-limb width of the rectangle), and the other side $C_{\odot}/4$ in length 
(the Earth-Sun
line-of-sight length of the rectangle), where $C_{\odot}$ is the solar circumference.  The area of this rectangular region, $C_{\odot}^2/(8*4)$,
is then $\sim$1/10 that of the surface area of the entire Sun.  Thus, our minimum (lower limit) estimate for the number of events 
observed by \citet{uritsky.et23}
over the entire Sun is 2300*10/360 per minute.  From Fig.\,5(a) of \citet{uritsky.et23} the mean lifetime of their observed events is 2.6 min, and
so the number of \citet{uritsky.et23} events seen at any one time on the entire Sun is $\sim$2300*10*2.6/360~$\sim$165.  As a minimum estimate
for the Earth-Sun line-of-sight length of the rectangle, we assume that the features observed by \citet{uritsky.et23} all come from only 
within $C_{\odot}/32$ of
the distance inside and beyond the limb, so that the line-of-sight side of the rectangle is now $C_{\odot}/16$ in length; this yields an upper limit 
for the estimate for the number of events at any one time on the entire Sun of $\sim$665.  Using these extrema for the range-bar limits and 
taking the value to
be the midpoint, 415 events at any given time over the entire Sun, we plot this value based on \citet{uritsky.et23} as the open-circle red entry 
on our Figure\,\ref{fourpt_dist_zu}.  We see that this value is close to our plotted value (solid black dot) for jetlets in Figure\,\ref{fourpt_dist_zu}.  

Thus, although we have had to make some assumptions in extrapolating the \citet{uritsky.et23} features
to the entire Sun, we can conclude that their observed features are broadly consistent with originating from jetlets.  Moreover, the
\citet{uritsky.et23} value provides an independent assessment for the number of outflowing features of the jetlet size scale, and 
this value is consistent with the value estimated from jetlet observations from \citet{raouafi.et23}.  Moreover, the two values from the
\citet{uritsky.et23} work and the \citet{raouafi.et23} work are both consistent with the size distribution of the number of erupting 
filament-like events on the Sun at any instant first presented in \citet{sterling.et16a}.

\section{Solar Wind Formation from Small-scale Eruptions that Make Jet-like Events}
\label{sec-solar_wind}

We can now propose how the solar wind might form from jet-like events, as proposed in \citet{moore.et11} and \citet{raouafi.et23}.
The latter work suggested that jetlets are the source of the solar wind.  Our discussion above in \S\ref{sec-distribution}, however,
suggests that jetlets are part of a continuous distribution of eruptive events, from large-scale eruptions that make CMEs, down
to spicules (or spicule-sized features).  Therefore, here we will describe the process as eruptions creating jetting events, where those events
could be jets, jetlets, or smaller features.

Figure~\ref{s_m20_matome_zu} summarizes different aspects of the process.  All of the jetting events would evolve 
as described in \S\ref{sec-jets}, and as shown in Figure~\ref{s_m20_matome_zu} panels~(a)---(c).  The magnetic flux
rope that holds the erupting minifilament (represented by the blue circle in the figure) would form via magnetic cancelation in 
the photosphere \citep[see fig.~4 in][]{panesar.et16a}.  This process continues until the flux rope becomes destabilized and 
erupts.  (Strictly speaking, a cool-material minifilament is not essential for this process. The miniature flux rope that forms could
erupt even without such cool material, or with very little cool material on it.  But the presence of the cool minifilament material allows
us to infer the presence of the erupting field in EUV images.)

Figure~\ref{s_m20_matome_zu} panels~(d)---(g) show a continuation of the evolution, with the external reconnection 
eroding away the entire outer envelope of the erupting minifilament.  The twist of the erupting minifilament flux rope is thereby
transmitted to the open field \citep{shibata.et86}, as mentioned in \S\ref{sec-jets}, and this twist can show up as swaying of the
white-light jets in coronagraph images \citep{moore.et15}.  This is an \al ic disturbance on the open field, and consequently it
propagates outward at the \al\ velocity.  As pointed out in \citet{sterling.et20a} this velocity decreases between the corona, where 
the \al\ velocity is $\sim$1000\,\kms, and the location of PSP; \citet{bale.et19} report it to be $\sim$100\,\kms\ at 36.6\,R$_{\odot}$.
This leads to a contraction of the \al\ pulse as it progresses outward, as indicated in panels~(h)---(j) of 
Figure~\ref{s_m20_matome_zu}.

Via this process, the material ejected from the Sun in the jetting event, specifically along the spire, can become solar wind material if
it gets out into the heliosphere (See \citeauthor{sterling.et20a}~\citeyear{sterling.et20a} for discussion of evidence that 
the jet material does indeed reach interplanetary space in some observed cases.)  At the very least, this can form some of the 
clumpy component (``flocculation") of the solar wind \citep[e.g.,][]{deforest.et16}.   Moreover, it can explain the presence of the 
``plasma jets" that are superimposed on the background Parker-like solar wind \citep{kasper.et19,raouafi.et23}, and also the correlation
between velocity microstreams and switchbacks \citep{neugebauer.et21}.

This also ties in the switchback structures with the solar wind in an intrinsic fashion: The jetting structures would produce the
clumpy component of the solar wind.  At the same time, the jetting structures would launch \al-wave pulses onto open field lines
that stretch into the heliosphere, and along those field lines those pulses would evolve into kinks in the field lines that would appear
as switchbacks.  

Switchbacks in the solar wind seem to cluster on preferential size scales about the size of supergranules \citep{bale.et21,fargette.et21},
and there is also evidence of switchbacks widths corresponding to that of photospheric granules \citep{fargette.et21}.
An average supergranule is about 30,000\,km across, and this size is within about a factor of four of the size of the minifilaments that 
erupt to make jets, with substantial variation in both values.  Similarly, microfilaments that might erupt to make some spicules are
expected to be within about a factor of three or four of that of a typical granule (size $\sim$1000\,km).  Therefore, jetting events 
preferentially of the sizes of coronal jets and of spicules could result in peaks in the observed widths of switchback clusters 
and individual switchbacks.  

One possibility for these switchback-width and switchback-cluster-width size scales might follow from the distribution of size scales on 
which magnetic cancelation occurs in the photosphere.  Convective motions in the photosphere occurs over a range of size scales, but 
with peaks on the granule and supergranule scales \citep{hathaway.et15}.  Magnetic fields cluster around the edges of granules and 
of supergranules, pushed there by the horizontal flows of those convective structures.  That could result in a preference for fields to 
shuffle around on those size scales, resulting in a preference for cancelations among opposite-polarity fields on those size scales.  
This could result in a preference for jetting events on these two size scales, meaning that coronal jets and spicule-sized jetting events 
are most common, resulting in the observed predominant width scales for switchbacks and switchback clusters (``patches").  It is unclear,
however, whether coronal jets are frequent enough for eruptions of near-supergranle-sized minifilaments to explain the observed
frequency of switchback patches.  Therefore, perhaps a more likely explanation for the these size scales is that most switchbacks 
result from small-scale eruptions that cause jetlet-sized and smaller jet-like features. Because both jetlets (see \S\ref{sec-jetlets}) and 
spicules \citep[e.g.,][]{samanta.et19} preferentially occur at network boundaries, switchbacks would also preferentially originate 
from those network boundaries; the network boundary is formed by supergranule motions, and so the network size reflects the
supergranule size, and this could explain why switchbacks could be nearly continuously launched in bunches of the size scale
of supergranules, resulting in the switchback patches on supergranule size scales.

\section{Discussion}
\label{sec-discussion}

We have shown that CMEs, coronal jets, jetlets, and spicule-sized features are consistent with all forming as a consequence of the
same basic mechanism on the Sun: Eruption of magnetic flux ropes.  The coronal jets and smaller features 
require the anemone setup, while it is not clear that large flares require a magnetic null over the erupting location; \citet{antiochos98}
argues that it is required, but other works suggest that it is not essential \citep{joshi.et17,jiang.et21}.  

In cases where an ejective eruption occurs inside of an anemone magnetic field configuration, one factor that determines 
whether a jet-like eruption occurs or a CME is ejected depends upon the how much of the erupting (mini)filament flux rope is 
eroded in the external reconnection with the surrounding open field.  In the jet case, the field is (essentially) completely eroded 
away, so that no flux rope escapes: everything that is expelled, including the hot plasma and the cool minifilament material,
escapes along the open coronal field.  In the case of the CME, the erupting filament flux rope and its magnetic envelope contains 
enough flux so that a remnant flux
rope survives the external reconnection, and that remnant flux rope escapes into the heliosphere and forms the core of the CME\@.  
(There are also cases, however, where a jet forms where the erupting minifilament flux rope survives but is non-ejective, with 
the erupting flux rope remaining confined near the base of the jet. This is discussed in \citeauthor{sterling.et22}~\citeyear{sterling.et22}.)

Thus, the features in Figure\,\ref{fourpt_dist_zu} that are small enough to survive as a jetting event rather than a CME are the ones
that might contribute to the solar wind.  \citet{raouafi.et23} suggests that jetlets are the main contributor to the solar wind, with the 
number of coronal jets being insufficient for the purpose.  Our work here says that jetting events on a variety of size scales are
available for contributing to the solar wind.  This is consistent 
with the evidence put forward by \citet{uritsky.et23}, who reached a similar conclusion through analysis of AIA images of outflows 
over six hours of continuous observations above a polar coronal 
hole. They found the mean size of the outflows, however, to be 3000---4000\,km (and this is consistent with the findings 
of \citeauthor{kumar.et23}~\citeyear{kumar.et23}), which might be described as large jetlets or 
small coronal jets, according to Figure\,\ref{fourpt_dist_zu}.  Furthermore, the most common-sized objects in their 
study (red open circle value in Fig.\,\ref{fourpt_dist_zu}) are near the size of our plotted jetlets point 
(second-fron-left black dot in Fig.\,\ref{fourpt_dist_zu}), and we deduce that the number of \citet{uritsky.et23} outflow events at
any given time over the entire Sun is nearly identical to our estimate of the same quantity for jetlets in Fig.\,\ref{fourpt_dist_zu}.  This
supports that the events observed by \citet{uritsky.et23} largely originate from jetlet-sized coronal-jet-like events.

Because the \al ic pulses imparted onto the open field would be longer close to the Sun, and shorter farther away 
from the Sun (Fig.~\ref{s_m20_matome_zu}(h---j)), on average the \al\ packets would appear as less kinked magnetic features
(or, smaller-rotation switchbacks) than after they travel farther from the Sun. As a result, they are less likely to be identified as
switchbacks near the Sun than farther from the Sun.  If features identified as switchbacks are not found (or more exactly: if only
very moderate magnetic field rotations are found) in the closest PSP
perihelions, then this will not necessarily imply that switchbacks originate only in solar wind farther from the Sun; it could instead be
that the seeds of those large-rotation switchbacks were launched by the \al ic pulses accompanying the jetting events, but those
pulses have yet to evolve into larger-rotation switchbacks as described in Figures~\ref{s_m20_matome_zu}(h---j).

It is unclear to us how the plugs of plasma expelled in the jetting events could make it into the solar wind and maintain the hot temperature
of solar-wind material, as adiabatic cooling would be expected if as the material disperses \citep{klimchuk12,sowmondal.et22}.  
One concept deserving of consideration is the consequences of the twists put onto the open flux tubes by the erupting minifilaments.

We envision many of the erupting minifilaments leading to jetting to have twist on them at the time of the eruption.  When this
twist gets transferred to open field, the twist becomes an \al ic pulse, as discussed above and in Figure~\ref{s_m20_matome_zu}.
These pulses will be of the form of a torsional \al ic twist, propagating outward along an open magnetic field.  \citet{hollweg.et82} found that 
in some cases, these propagating \al\ waves can contribute to heating of the plasma through which they propagate.  
They found that as these \al\ waves propagate up into the atmosphere, they can nonlinearly 
couple to fast- and slow-mode wave modes, which are compressive and can steepen into shocks and impart heating.   They found
much of 
this steepening to occur in the chromosphere, where the magnetic flux tube undergoes rapid expansion, leading to a density drop and increase in 
the \al\ speed.  This inspired much work trying to connect these waves with spicule production \citep{hollweg.et82,kudoh.et99,matsumoto.et10}.

For the minifilament-eruption jetting mechanism, whether twist will be imparted onto the open field depends on the
size of the jet-like event.  In Figure~\ref{fourpt_dist_zu}, only the smallest-size-scale events, representing the left-most point (smallest-sized
erupting filament-like features, leading to spicule-sized events) and not to the second point (representing the jetlets) would have 
external reconnections in the chromosphere, and thus be subject to the severe nonlinear effects discussed in \citet{hollweg.et82}.  These 
cases, however, are the most numerous, and therefore for them there is a valid question of whether this wave-mode coupling can 
result in plasma heating.  A recent numerical simulation \citep{soler.et19} using a train of such \al\ waves and diffusive processes in the chromosphere
(ohmic magnetic diffusion, or ambipolar diffusion, \citeauthor{khomenko.et12}~\citeyear{khomenko.et12}; and ion-neutral collisions),
found only a small fraction of the energy to reach the corona; $\sim$10$^5$\,erg\,cm$^{-2}$\,s$^{-1}$, but this might contribute
to such heating.   Other work, however, shows that the specific contribution of the \al\ waves to heating of the solar atmosphere is 
dependent upon the specific parameters of the flux tube and wave-launch conditions \citep{antolin.et10}.  It would be of interest to 
see simulations representing the consequences of twists imparted in a manner representing the processes in Figure~\ref{s_m20_matome_zu}, and extending out to the location of PSP observations, to see whether heating via 
\al\ wave-mode coupling to slow and fast modes leads to shocks that can heat the local solar wind plasma.

\citet{karpen.et17} undertook a 3D MHD simulation of a jet in a coronal-hole
field with an embedded bipole at the base of their calculation region.  Their setup is similar to the coronal-jet scenario presented in 
\citet{sterling.et15} and discussed in \S\ref{sec-jets}, with a minority magnetic polarity flux concentration surrounded by a broad region of 
majority flux, and a null point in the corona above the minority polarity flux.  An important difference, though, is that they assumed 
symmetry to the system; in contrast the \citet{sterling.et15}  geometry is asymmetric, with one part of the anemone in a non-potential 
state and holding a minifilament, while the rest of the anemone is 
roughly potential (see Fig.\,\ref{s_m20_matome_zu}(a)).  In the \citet{karpen.et17} case, a jet results when a symmetric subsonic twist is 
imparted to the base of the setup, resulting in a puffing out of the anemone base field, until a kink instability sets in resulting in 
a jet traveling outward as a twist wave along the coronal field extending radially outward above the anemone's null; this is different
from the asymmetric minifilament/flux rope eruption deduced from our observations (Figs.\,\ref{s_m20_matome_zu}(b) and (c)).  

In followup studies to \citet{karpen.et17}, \citet{uritsky.et17} examined the structure and turbulent dynamics of the \citet{karpen.et17} simulated jet as it
propagates out into the heliosphere, and \citet{roberts.et18} calculate predictions for what PSP will see from the resulting 
outward-propagating simulated jet.  \citet{uritsky.et17} calculated that there would be a set of dynamic regions behind the
outward-propagating leading edge of the jet, involving turbulent structures on various size scales \citep[see Fig.\,14 of][]{uritsky.et17}.
While these are interesting predictions, the manner in which the simulated jet is initiated differs from what we infer for jet onset
based on our observations; our observations imply that the jet starts with the eruption of a minifilament/flux rope from an
asymmetric magnetic anemone at the base of the jet-spire field.  It would be of interest if the analysis of \citet{uritsky.et17} and \citet{roberts.et18}
could be carried out in a geometry that mimics more closely these observations, such as the simulation geometry of \citet{wyper.et17}, 
to confirm whether the same far-from-Sun features develop.  A further refinement would be to incorporate magnetic flux cancelation as
the process for initiating the flux-rope eruption in a simulation with a coronal topology such as that of \citet{wyper.et17}.

Returning to the issue of observations of jets on various size scales, there is the possibility that it may be difficult to count 
smaller jets using imaging data alone.  This is exemplified by two features, called ``dark jets" and ``inconspicuous jets."

\citet{young15} observed an on-disk coronal hole using spectral scans with the EUV Imaging Spectrometer (EIS) on \hinode. Doppler
velocity maps of the region in the 195.12\,\AA\ \fexii\ line revealed numerous transient localized regions of blue-shifted 
upflows.  Corresponding observations in AIA\,193\,\AA\ images showed only either a weak counterpart or no signature at all at the 
upflow locations.  He called these EIS features {\it dark jets}, and found them to be as common as regular coronal hole jets, but
with an intensity in AIA so low that the dark jets must have a mass flux one or two magnitudes lower.

In a similar fashion, \citet{schwanitz.et21} also looked at EIS spectral scans.  They focused on 14 Doppler localized, transient 
EIS upflow regions, and looked for counterparts in AIA EUV images, and SXT images for the five of the 14 events 
observed with \hinode/XRT\@.  They classified only one of of the events as 
``obvious jets” and one as a ``bright point with jet.”  They classified seven as ``small-scale brightenings/eruptions,” three 
as ``bright points,” and two as ``unclear.”  Four of the 14 were at low latitude, allowing for comparison with HMI magnetograms. 
They report that three of these were ``bright point" events and all three of these showed evidence for flux 
cancelation, and they say that the fourth event was unclear and showed no HMI feature.

\citet{sterling.et22a} looked more closely at the five \citet{schwanitz.et21} events having XRT data.   One of those 
five was the one categorized as ``obvious jets" in \citet{schwanitz.et21}, but the other four were not classified 
as jets.  But upon closely comparing dynamic motions in AIA and XRT images, \citet{sterling.et22a} concluded that all 
five were consistent with being coronal jets that were very inconspicuous in the images.  The evidence for this includes 
that in all five events they found evidence for eruption 
of a cool minifilament coinciding with the EIS upflow locations, and the sizes and near-eruption-onset-time speeds of those
erupting minifilaments fall into the ranges of values for the same parameters found from confirmed coronal jets in previous
studies \citep{sterling.et15,sterling.et22}.  These erupting minifilaments were either confined to the base location, or perhaps 
ejective but where the cool material become extremely tenuous low down so that they are not obvious beyond a small-ish height;
this perhaps explains why that material was not detected as a jet in  \citet{schwanitz.et21}.  Also, upon close inspection (including
using difference images in one case), all five events showed a jet spire, and in four of the five cases where spire motion could be
detected, that spire moved away from the JBP with time, which is also consistent with confirmed coronal jets \citep{baikie.et22}.
All five of the \citet{sterling.et22a} events were near the north polar region.  One of the five events however occurred at low enough 
latitude to allow comparisons with HMI, and those magnetograms were consistent with flux cancelation triggering that event.

The \citet{young15} ``dark jets" name applies to whatever it is that makes the EIS upflows, independent of whether they are true coronal 
jets.  The ``inconspicuous jets" described by \citet{sterling.et22a}, on the other hand, are observed features that are strongly
consistent with being true coronal jets, but are hard to detect due to their weak intensity in EUV and SXRs.  Whether the 
\citet{young15} dark jets are inconspicuous jets is not known at this time.  One difference is the wavelength coverage inspected:
\citet{young15} investigated only one AIA channel (193\,\AA), while \citet{sterling.et22a} used four AIA channels (171, 193, 211,
and 304\,\AA) and SXRs for their investigations.  Therefore, it is unknown whether the dark jets are truly coronal jets.  Similarly,
it is unknown whether the remaining eight events of \citet{schwanitz.et21} (the nine other than the five studied by  
\citeauthor{sterling.et22a}~\citeyear{sterling.et22a}, and the one additional upflow location characterized as a ``bright point 
with jet" in \citeauthor{schwanitz.et21}~\citeyear{schwanitz.et21}) are inconspicuous jets.  But the  \citet{sterling.et22a} 
study implies that the coronal-jet mechanism is the cause of some upflows that are readily seen in EIS Doppler 
spectral scans but that are difficult to detect in EUV and SXR images.  A key point is that this mechanism is 
responsible for more than just the jets that are obviously and easily detectable in AIA and EUV images.  All five 
of the events studied in \citet{sterling.et22a} \citep[which were selected from EIS Doppler data in][]{schwanitz.et21}  
would have been too weak and feeble in XRT and/or EUV images to be selected as examples of jets in our previous 
jet studies \citep[e.g.,][]{moore.et10,moore.et13,sterling.et15,panesar.et16a,panesar.et18a,mcglasson.et19,sterling.et22}.

\vspace{1cm}

\comment{

} 

\begin{acknowledgments}
ACS thanks N. Raouafi for stimulating discussions.  We thank an anonymous referee for asking for more discussion of the observations
of \citet{uritsky.et23}, and for providing other helpful comments.  A.C.S., R.L.M. and N.K.P. received funding from the Heliophysics Division of NASA's Science 
Mission Directorate through the Heliophysics Supporting Research (HSR, grant No.~20-HSR20\_2-0124) Program, 
and the Heliophysics Guest Investigators program.  ACS and RLM were also supported 
through the Heliophysics System Observatory Connect (HSOC, grant No.~80NSSC20K1285) 
Program.  N.K.P. received additional support through a NASA \sdo/AIA grant.   
We acknowledge the use of AIA data. AIA is an instrument onboard \sdo, a mission of
NASA's Living With a Star program.
\end{acknowledgments}

\bibliography{ms}

\begin{thebibliography}{}
\expandafter\ifx\csname natexlab\endcsname\relax\def\natexlab#1{#1}\fi
\providecommand{\url}[1]{\href{#1}{#1}}
\providecommand{\dodoi}[1]{doi:~\href{http://doi.org/#1}{\nolinkurl{#1}}}
\providecommand{\doeprint}[1]{\href{http://ascl.net/#1}{\nolinkurl{http://ascl.net/#1}}}
\providecommand{\doarXiv}[1]{\href{https://arxiv.org/abs/#1}{\nolinkurl{https://arxiv.org/abs/#1}}}

\bibitem[{Adams {et~al.}(2014)Adams, Sterling, Moore, \& Gary}]{adams.et14}
Adams, M., Sterling, A.~C., Moore, R.~L., \& Gary, G.~A. 2014, Astrophysical
  Journal, 783, 11, \dodoi{10.1088/0004-637X/783/1/11}

\bibitem[{Antiochos(1998)}]{antiochos98}
Antiochos, S.~K. 1998, Astrophysical Journal, 502, L181, \dodoi{10.1086/311507}

\bibitem[{{Antolin} \& {Shibata}(2010)}]{antolin.et10}
{Antolin}, P., \& {Shibata}, K. 2010, \apj, 712, 494,
  \dodoi{10.1088/0004-637X/712/1/494}

\bibitem[{{Baikie} {et~al.}(2022){Baikie}, {Sterling}, {Moore}, {Alexander},
  {Falconer}, {Savcheva}, \& {Savage}}]{baikie.et22}
{Baikie}, T.~K., {Sterling}, A.~C., {Moore}, R.~L., {et~al.} 2022, arXiv
  e-prints, arXiv:2201.08882.
\newblock \doarXiv{2201.08882}

\bibitem[{Bale {et~al.}(2019)Bale, Badman, Bonnell, Bowen, Burgess, Case,
  Cattell, Chandran, Chaston, Chen, Drake, Dudok~de Witt, Eastwood, Ergun,
  Farrell, Fong, Goetz, Goldstein, Goodrich, Harvey, Horbury, Howes, Kasper,
  Kellogg, Klimcuk, Korreck, Krasnoselskikh, Krucker, Laker, Larson, MacDowall,
  Maksimovic, Malaspina, Martinez-Oliveros, McComas, Meyer-Vernet, Moncuquet,
  Mozer, Phan, Pulupa, Raouafi, Salem, Stansby, Stevens, Szabo, Velli, Woolley,
  \& Wygant}]{bale.et19}
Bale, S.~D., Badman, S.~T., Bonnell, J.~W., {et~al.} 2019, Nature, 576, 237,
  \dodoi{10.1038/s41586-019-1818-7}

\bibitem[{{Bale} {et~al.}(2021){Bale}, {Horbury}, {Velli}, {Desai}, {Halekas},
  {McManus}, {Panasenco}, {Badman}, {Bowen}, {Chandran}, {Drake}, {Kasper},
  {Laker}, {Mallet}, {Matteini}, {Phan}, {Raouafi}, {Squire}, {Woodham}, \&
  {Woolley}}]{bale.et21}
{Bale}, S.~D., {Horbury}, T.~S., {Velli}, M., {et~al.} 2021, \apj, 923, 174,
  \dodoi{10.3847/1538-4357/ac2d8c}

\bibitem[{{Berghmans} {et~al.}(2021){Berghmans}, {Auch{\`e}re}, {Long},
  {Soubri{\'e}}, {Mierla}, {Zhukov}, {Sch{\"u}hle}, {Antolin}, {Harra},
  {Parenti}, {Podladchikova}, {Aznar Cuadrado}, {Buchlin}, {Dolla}, {Verbeeck},
  {Gissot}, {Teriaca}, {Haberreiter}, {Katsiyannis}, {Rodriguez}, {Kraaikamp},
  {Smith}, {Stegen}, {Rochus}, {Halain}, {Jacques}, {Thompson}, \&
  {Inhester}}]{berghmans.et21}
{Berghmans}, D., {Auch{\`e}re}, F., {Long}, D.~M., {et~al.} 2021, \aap, 656,
  L4, \dodoi{10.1051/0004-6361/202140380}

\bibitem[{{Chen} {et~al.}(2020){Chen}, {Hong}, {Yang}, {Xu}, \&
  {Yang}}]{chen.et20}
{Chen}, H., {Hong}, J., {Yang}, B., {Xu}, Z., \& {Yang}, J. 2020, \apj, 902, 8,
  \dodoi{10.3847/1538-4357/abb1c1}

\bibitem[{Chen {et~al.}(2012)Chen, Zhang, \& Ma}]{chen.et12}
Chen, H., Zhang, J., \& Ma, S. 2012, Research in Astronomy and Astrophysics,
  12, 573, \dodoi{10.1088/1674-4527/12/5/009}

\bibitem[{{Chen}(2011)}]{chen11}
{Chen}, P.~F. 2011, Living Reviews in Solar Physics, 8, 1,
  \dodoi{10.12942/lrsp-2011-1}

\bibitem[{{Chitta} {et~al.}(2023){Chitta}, {Zhukov}, {Berghmans}, {Peter},
  {Parenti}, {Mandal}, {Aznar Cuadrado}, {Sch{\"u}hle}, {Teriaca},
  {Auch{\`e}re}, {Barczynski}, {Buchlin}, {Harra}, {Kraaikamp}, {Long},
  {Rodriguez}, {Schwanitz}, {Smith}, {Verbeeck}, \& {Seaton}}]{chitta.et23}
{Chitta}, L.~P., {Zhukov}, A.~N., {Berghmans}, D., {et~al.} 2023, Science, 381,
  867, \dodoi{10.1126/science.ade5801}

\bibitem[{Curdt {et~al.}(2012)Curdt, Tian, \& Kamio}]{curdt.et12}
Curdt, W., Tian, H., \& Kamio, S. 2012, Solar Physics, 280, 417,
  \dodoi{10.1007/s11207-012-9940-9}

\bibitem[{{DeForest} {et~al.}(2016){DeForest}, {Matthaeus}, {Viall}, \&
  {Cranmer}}]{deforest.et16}
{DeForest}, C.~E., {Matthaeus}, W.~H., {Viall}, N.~M., \& {Cranmer}, S.~R.
  2016, \apj, 828, 66, \dodoi{10.3847/0004-637X/828/2/66}

\bibitem[{{Fargette} {et~al.}(2021){Fargette}, {Lavraud}, {Rouillard},
  {R{\'e}ville}, {Dudok De Wit}, {Froment}, {Halekas}, {Phan}, {Malaspina},
  {Bale}, {Kasper}, {Louarn}, {Case}, {Korreck}, {Larson}, {Pulupa}, {Stevens},
  {Whittlesey}, \& {Berthomier}}]{fargette.et21}
{Fargette}, N., {Lavraud}, B., {Rouillard}, A.~P., {et~al.} 2021, \apj, 919,
  96, \dodoi{10.3847/1538-4357/ac1112}

\bibitem[{{Hathaway} {et~al.}(2015){Hathaway}, {Teil}, {Norton}, \&
  {Kitiashvili}}]{hathaway.et15}
{Hathaway}, D.~H., {Teil}, T., {Norton}, A.~A., \& {Kitiashvili}, I. 2015,
  \apj, 811, 105, \dodoi{10.1088/0004-637X/811/2/105}

\bibitem[{{Hinode Review Team} {et~al.}(2019){Hinode Review Team}, {Khalid},
  {Patrick}, {Baker}, R., \& {et al.}}]{hinode.et19}
{Hinode Review Team}, {Khalid}, A.-J., {Patrick}, A., {et~al.} 2019,
  Publications of the Astronomical Society of Japan, 71, id.R1,
  \dodoi{10.1093/pasj/psz084}

\bibitem[{Hollweg {et~al.}(1982)Hollweg, Jackson, \& Galloway}]{hollweg.et82}
Hollweg, J.~V., Jackson, S., \& Galloway, D. 1982, Solar Physics, 75, 35,
  \dodoi{10.1007/BF00153458}

\bibitem[{Hong {et~al.}(2013)Hong, Jiang, Yang, Zheng, Bi, Li, Yang, \&
  Yang}]{hong.et13}
Hong, J., Jiang, Y., Yang, J., {et~al.} 2013, Research in Astronomy and
  Astrophysics, 13, 253, \dodoi{10.1088/1674-4527/13/3/001}

\bibitem[{Iijima \& Yokoyama(2017)}]{iijima.et17}
Iijima, H., \& Yokoyama, T. 2017, Astrophysical Journal, 848, 38,
  \dodoi{10.3847/1538-4357/aa8ad1}

\bibitem[{{Jiang} {et~al.}(2021){Jiang}, {Feng}, {Liu}, {Yan}, {Hu}, {Moore},
  {Duan}, {Cui}, {Zuo}, {Wang}, \& {Wei}}]{jiang.et21}
{Jiang}, C., {Feng}, X., {Liu}, R., {et~al.} 2021, Nature Astronomy, 5, 1126,
  \dodoi{10.1038/s41550-021-01414-z}

\bibitem[{Joshi {et~al.}(2017)Joshi, Thalmann, Mitra, Chandra, \&
  Veronig}]{joshi.et17}
Joshi, B., Thalmann, J.~K., Mitra, P.~K., Chandra, R., \& Veronig, A.~M. 2017,
  Astrophysical Journal, 851, 29, \dodoi{10.3847/1538-4357/aa9564}

\bibitem[{Joshi {et~al.}(2018)Joshi, Nishizuka, Filippov, Magara, \&
  Tlatov}]{joshi.et18}
Joshi, N.~C., Nishizuka, N., Filippov, B., Magara, T., \& Tlatov, A.~G. 2018,
  MNRAS, 476, 1286, \dodoi{10.1093/mnras/sty322}

\bibitem[{Kamio {et~al.}(2010)Kamio, Curdt, Teriaca, Inhester, \&
  Solanki}]{kamio.et10}
Kamio, S., Curdt, W., Teriaca, L., Inhester, B., \& Solanki, S.~K. 2010,
  Astronomy and Astrophysics, 510, 1, \dodoi{10.1051/0004-6361/200913269}

\bibitem[{{Karpen} {et~al.}(2017){Karpen}, {DeVore}, {Antiochos}, \&
  {Pariat}}]{karpen.et17}
{Karpen}, J.~T., {DeVore}, C.~R., {Antiochos}, S.~K., \& {Pariat}, E. 2017,
  \apj, 834, 62, \dodoi{10.3847/1538-4357/834/1/62}

\bibitem[{Kasper {et~al.}(2019)Kasper, Bale, Belcher, Berthomier, Case,
  Chandran, Curtis, Gallagher, Gary, Golub, Halekas, Ho, Horbury, Hu, Huang,
  Klein, Korreck, Larson, Livi, Maruca, Lavraud, Louarn, Maksimovic,
  Martinovic, McGinnis, Pogorelov, Richardson, Skoug, Steinberg, Stevens,
  Szabo, Velli, Whittlesey, Wright, Zank, MacDowall, McComas, McNutt, Pulupa,
  Raouafi, \& Schwadron}]{kasper.et19}
Kasper, J.~C., Bale, S.~D., Belcher, J.~W., {et~al.} 2019, Nature, 576, 228,
  \dodoi{10.1038/s41586-019-1813-z}

\bibitem[{{Khomenko} \& {Collados}(2012)}]{khomenko.et12}
{Khomenko}, E., \& {Collados}, M. 2012, \apj, 747, 87,
  \dodoi{10.1088/0004-637X/747/2/87}

\bibitem[{Klimchuk(2012)}]{klimchuk12}
Klimchuk, J.~A. 2012, Journal of Geophysical Research, 117, A12102,
  \dodoi{10.1029/2012JA018170}

\bibitem[{{Kotani} \& {Shibata}(2020)}]{kotani.et20}
{Kotani}, Y., \& {Shibata}, K. 2020, \pasj, 72, 75,
  \dodoi{10.1093/pasj/psaa064}

\bibitem[{Kudoh \& Shibata(1999)}]{kudoh.et99}
Kudoh, T., \& Shibata, K. 1999, Astrophysical Journal, 514, 493,
  \dodoi{10.1086/306930}

\bibitem[{{Kumar} {et~al.}(2023){Kumar}, {Karpen}, {Uritsky}, {Deforest},
  {Raouafi}, {DeVore}, \& {Antiochos}}]{kumar.et23}
{Kumar}, P., {Karpen}, J.~T., {Uritsky}, V.~M., {et~al.} 2023, \apjl, 951, L15,
  \dodoi{10.3847/2041-8213/acd54e}

\bibitem[{{Kumar} {et~al.}(2022){Kumar}, {Karpen}, {Uritsky}, {Deforest},
  {Raouafi}, \& {Richard DeVore}}]{kumar.et22}
---. 2022, \apj, 933, 21, \dodoi{10.3847/1538-4357/ac6c24}

\bibitem[{Liu {et~al.}(2019)Liu, Wang, \& Erd{\'e}lyi}]{liu.et19}
Liu, J., Wang, Y., \& Erd{\'e}lyi, R. 2019, Frontiers in Astronomy and Space
  Sciences, 6, 44L, \dodoi{10.3389/fspas.2019.00044}

\bibitem[{Mart{\'i}nez-Sykora {et~al.}(2017)Mart{\'i}nez-Sykora, De~Pontieu,
  Hansteen, Rouppe van~der Voort, Carlsson, \& Pereira}]{martinez-sykora.et17}
Mart{\'i}nez-Sykora, J., De~Pontieu, B., Hansteen, V.~H., {et~al.} 2017,
  Science, 356, 1269, \dodoi{10.1126/science.aah5412}

\bibitem[{{Matsumoto} \& {Shibata}(2010)}]{matsumoto.et10}
{Matsumoto}, T., \& {Shibata}, K. 2010, \apj, 710, 1857,
  \dodoi{10.1088/0004-637X/710/2/1857}

\bibitem[{McGlasson {et~al.}(2019)McGlasson, Panesar, Sterling, \&
  Moore}]{mcglasson.et19}
McGlasson, R.~A., Panesar, N.~K., Sterling, A.~C., \& Moore, R.~L. 2019, \apj,
  882, 16, \dodoi{10.3847/1538-4357/ab2fe3}

\bibitem[{Moore {et~al.}(2010)Moore, Cirtain, Sterling, \&
  Falconer}]{moore.et10}
Moore, R.~L., Cirtain, J.~W., Sterling, A.~C., \& Falconer, D.~A. 2010,
  Astrophysical Journal, 720, 757, \dodoi{10.1088/0004-637X/720/1/757}

\bibitem[{Moore {et~al.}(2011)Moore, Sterling, Cirtain, \&
  Falconer}]{moore.et11}
Moore, R.~L., Sterling, A.~C., Cirtain, J.~W., \& Falconer, D.~A. 2011,
  Astrophysical Journal, 731L, 18, \dodoi{10.1088/2041-8205/731/1/L18}

\bibitem[{Moore {et~al.}(2013)Moore, Sterling, Falconer, \& Robe}]{moore.et13}
Moore, R.~L., Sterling, A.~C., Falconer, D.~A., \& Robe, D. 2013, Astrophysical
  Journal, 769, 134, \dodoi{10.1088/0004-637X/769/2/134}

\bibitem[{Moore {et~al.}(2015)Moore, Sterling, \& Falconer}]{moore.et15}
Moore, R.~L., Sterling, R.~L., \& Falconer, D.~A. 2015, Astrophysical Journal,
  806, 11, \dodoi{10.1088/0004-637X/806/1/11}

\bibitem[{Moore {et~al.}(2018)Moore, Sterling, \& Panesar}]{moore.et18}
Moore, R.~L., Sterling, R.~L., \& Panesar, N.~K. 2018, Astrophysical Journal,
  859, 3, \dodoi{10.3847/1538-4357/aabe79}

\bibitem[{{Morton} {et~al.}(2012){Morton}, {Srivastava}, \&
  {Erd{\'e}lyi}}]{morton.et12}
{Morton}, R.~J., {Srivastava}, A.~K., \& {Erd{\'e}lyi}, R. 2012, \aap, 542,
  A70, \dodoi{10.1051/0004-6361/201117218}

\bibitem[{{Muglach}(2021)}]{muglach21}
{Muglach}, K. 2021, \apj, 909, 133, \dodoi{10.3847/1538-4357/abd5ad}

\bibitem[{{Neugebauer} \& {Sterling}(2021)}]{neugebauer.et21}
{Neugebauer}, M., \& {Sterling}, A.~C. 2021, \apjl, 920, L31,
  \dodoi{10.3847/2041-8213/ac2945}

\bibitem[{Nistic{\`o} {et~al.}(2009)Nistic{\`o}, Bothmer, Patsourakos, \&
  Zimbardo}]{nistico.et09}
Nistic{\`o}, G., Bothmer, V., Patsourakos, S., \& Zimbardo, G. 2009, Solar
  Physics, 259, 87, \dodoi{10.1007/s11207-009-9424-8}

\bibitem[{Panesar {et~al.}(2016{\natexlab{a}})Panesar, Sterling, \&
  Moore}]{panesar.et16b}
Panesar, N.~K., Sterling, A.~C., \& Moore, R.~L. 2016{\natexlab{a}},
  Astrophysical Journal, 822L, 7, \dodoi{10.3847/2041-8205/822/2/L23}

\bibitem[{Panesar {et~al.}(2017)Panesar, Sterling, \& Moore}]{panesar.et17}
---. 2017, Astrophysical Journal, 844, 131, \dodoi{10.3847/1538-4357/aa7b77}

\bibitem[{Panesar {et~al.}(2018{\natexlab{a}})Panesar, Sterling, \&
  Moore}]{panesar.et18a}
---. 2018{\natexlab{a}}, Astrophysical Journal, 853, 189,
  \dodoi{10.3847/1538-4357/aaa3e9}

\bibitem[{Panesar {et~al.}(2016{\natexlab{b}})Panesar, Sterling, Moore, \&
  Chakrapani}]{panesar.et16a}
Panesar, N.~K., Sterling, A.~C., Moore, R.~L., \& Chakrapani, P.
  2016{\natexlab{b}}, Astrophysical Journal, 832L, 7,
  \dodoi{10.3847/2041-8205/832/1/L7}

\bibitem[{Panesar {et~al.}(2018{\natexlab{b}})Panesar, Sterling, Moore, Tiwari,
  De~Pontieu, \& Norton}]{panesar.et18b}
Panesar, N.~K., Sterling, A.~C., Moore, R.~L., {et~al.} 2018{\natexlab{b}},
  Astrophysical Journal, 868L, 27, \dodoi{10.3847/2041-8213/aaef37}

\bibitem[{{Panesar} {et~al.}(2021){Panesar}, {Tiwari}, {Berghmans}, {Cheung},
  {M{\"u}ller}, {Auchere}, \& {Zhukov}}]{panesar.et21}
{Panesar}, N.~K., {Tiwari}, S.~K., {Berghmans}, D., {et~al.} 2021, \apjl, 921,
  L20, \dodoi{10.3847/2041-8213/ac3007}

\bibitem[{{Panesar} {et~al.}(2022){Panesar}, {Tiwari}, {Moore}, {Sterling}, \&
  {De Pontieu}}]{panesar.et22}
{Panesar}, N.~K., {Tiwari}, S.~K., {Moore}, R.~L., {Sterling}, A.~C., \& {De
  Pontieu}, B. 2022, \apj, 939, 25, \dodoi{10.3847/1538-4357/ac8d65}

\bibitem[{{Panesar} {et~al.}(2019){Panesar}, {Sterling}, {Moore}, {Winebarger},
  {Tiwari}, {Savage}, {Golub}, {Rachmeler}, {Kobayashi}, {Brooks}, {Cirtain},
  {De Pontieu}, {McKenzie}, {Morton}, {Peter}, {Testa}, {Walsh}, \&
  {Warren}}]{panesar.et19}
{Panesar}, N.~K., {Sterling}, A.~C., {Moore}, R.~L., {et~al.} 2019, \apjl, 887,
  L8, \dodoi{10.3847/2041-8213/ab594a}

\bibitem[{{Paraschiv} \& {Donea}(2019)}]{paraschiv.et19}
{Paraschiv}, A.~R., \& {Donea}, A. 2019, \apj, 873, 110,
  \dodoi{10.3847/1538-4357/ab04a6}

\bibitem[{{Paraschiv} {et~al.}(2020){Paraschiv}, {Donea}, \&
  {Leka}}]{paraschiv.et20}
{Paraschiv}, A.~R., {Donea}, A., \& {Leka}, K.~D. 2020, \apj, 891, 149,
  \dodoi{10.3847/1538-4357/ab7246}

\bibitem[{{Paraschiv} {et~al.}(2022){Paraschiv}, {Donea}, \&
  {Judge}}]{paraschiv.et22}
{Paraschiv}, A.~R., {Donea}, A.~C., \& {Judge}, P.~G. 2022, \apj, 935, 172,
  \dodoi{10.3847/1538-4357/ac80fb}

\bibitem[{Patsourakos {et~al.}(2008)Patsourakos, Pariat, Vourlidas, Antiochos,
  \& Wuelser}]{patsourakos.et08}
Patsourakos, S., Pariat, E., Vourlidas, A., Antiochos, S.~K., \& Wuelser, J.~P.
  2008, Astrophysical Journal, 680, L73, \dodoi{10.1086/589769}

\bibitem[{Pike \& Mason(1998)}]{pike.et98}
Pike, C.~D., \& Mason, H.~E. 1998, Solar Physics, 182, 333,
  \dodoi{10.1023/A:1005065704108}

\bibitem[{Rachmeler(2019)}]{rachmeler.et19}
Rachmeler, L. A. e.~a. 2019

\bibitem[{Raouafi {et~al.}(2010)Raouafi, Georgoulis, Rust, \&
  Bernasconi}]{raouafi.et10}
Raouafi, N.-E., Georgoulis, M.~K., Rust, D.~M., \& Bernasconi, P.~N. 2010,
  Astrophysical Journal, 718, 981, \dodoi{10.1088/0004-637X/718/2/981}

\bibitem[{Raouafi \& Stenborg(2014)}]{raouafi.et14}
Raouafi, N.~E., \& Stenborg, G. 2014, Astrophysical Journal, 787, 118,
  \dodoi{10.1088/0004-637X/787/2/118}

\bibitem[{Raouafi {et~al.}(2016)Raouafi, Patsourakos, Pariat, Young, Sterling,
  Savcheva, Shimojo, Moreno-Insertis, DeVore, Archontis, Török, Mason, Curdt,
  Meyer, Dalmasse, \& Matsui}]{raouafi.et16}
Raouafi, N.~E., Patsourakos, S., Pariat, E., {et~al.} 2016, Space Science
  Reviews, 201, 1, \dodoi{10.1007/s11214-016-0260-5}

\bibitem[{{Raouafi} {et~al.}(2023){Raouafi}, {Stenborg}, {Seaton}, {Wang},
  {Wang}, {DeForest}, {Bale}, {Drake}, {Uritsky}, {Karpen}, {DeVore},
  {Sterling}, {Horbury}, {Harra}, {Bourouaine}, {Kasper}, {Kumar}, {Phan}, \&
  {Velli}}]{raouafi.et23}
{Raouafi}, N.~E., {Stenborg}, G., {Seaton}, D.~B., {et~al.} 2023, \apj, 945,
  28, \dodoi{10.3847/1538-4357/acaf6c}

\bibitem[{Roberts {et~al.}(2018)Roberts, Uritsky, DeVore, \&
  Karpen}]{roberts.et18}
Roberts, M.~A., Uritsky, V.~M., DeVore, C.~R., \& Karpen, J.~T. 2018,
  Astrophysical Journal, 866, 14, \dodoi{10.3847/1538-4357/aadb41}

\bibitem[{Samanta {et~al.}(2019)Samanta, Tian, Yurchyshyn, Peter, Cao,
  Sterling, Erd{'e}lyi, Ahn, Feng, Utz, Banerjee, \& Chen}]{samanta.et19}
Samanta, T., Tian, H., Yurchyshyn, V., {et~al.} 2019, Science, 366, 890,
  \dodoi{10.1126/science.aaw2796}

\bibitem[{Savcheva {et~al.}(2007)Savcheva, Cirtain, Deluca, Lundquist, Golub,
  Weber, Shimojo, Shibasaki, Sakao, Narukage, Tsuneta, \& Kano}]{savcheva.et07}
Savcheva, A., Cirtain, J., Deluca, E.~E., {et~al.} 2007, Publications of the
  Astronomical Society of Japan, 59, 771, \dodoi{10.1093/pasj/59.sp3.S771}

\bibitem[{{Schmieder}(2022)}]{schmieder22}
{Schmieder}, B. 2022, arXiv e-prints, arXiv:2201.11541.
\newblock \doarXiv{2201.11541}

\bibitem[{{Schwanitz} {et~al.}(2021){Schwanitz}, {Harra}, {Raouafi},
  {Sterling}, {Moreno Vacas}, {del Toro Iniesta}, {Orozco Su{\'a}rez}, \&
  {Hara}}]{schwanitz.et21}
{Schwanitz}, C., {Harra}, L., {Raouafi}, N.~E., {et~al.} 2021, \solphys, 296,
  175, \dodoi{10.1007/s11207-021-01915-0}

\bibitem[{{Shen}(2021)}]{shen21}
{Shen}, Y. 2021, Proceedings of the Royal Society of London Series A, 477, 217,
  \dodoi{10.1098/rspa.2020.0217}

\bibitem[{Shen {et~al.}(2012)Shen, Liu, Su, \& Deng}]{shen.et12}
Shen, Y., Liu, Y., Su, J., \& Deng, Y. 2012, Astrophysical Journal, 745, 164,
  \dodoi{10.1088/0004-637X/745/2/164}

\bibitem[{Shibata \& Magara(2011)}]{shibata.et11}
Shibata, K., \& Magara, T. 2011, LRSP, 8, 6

\bibitem[{Shibata \& Uchida(1986)}]{shibata.et86}
Shibata, K., \& Uchida, Y. 1986, Solar Physics, 178, 379

\bibitem[{Shibata {et~al.}(1992)Shibata, Ishido, Acton, Strong, Hirayama,
  Uchida, McAllister, Matsumoto, Tsuneta, Shimizu, Hara, Sakurai, Ichimoto,
  Nishino, \& Ogawara}]{shibata.et92}
Shibata, K., Ishido, Y., Acton, L.~W., {et~al.} 1992, Publications of the
  Astronomical Society of Japan, 44, L173

\bibitem[{Shibata {et~al.}(2007)Shibata, Nakamura, Matsumoto, Otsuji, Okamoto,
  Nishizuka, Kawate, Watanabe, Nagata, UeNo, Kitai, Nozawa, Tsuneta, Suematsu,
  Ichimoto, Shimizu, Katsukawa, Tarbell, Berger, Lites, Shine, \&
  Title}]{shibata.et07}
Shibata, K., Nakamura, T., Matsumoto, T., {et~al.} 2007, Science, 318, 1591,
  \dodoi{10.1126/science.1146708}

\bibitem[{Shimojo \& Shibata(2000)}]{shimojo.et00}
Shimojo, M., \& Shibata, K. 2000, Astrophysical Journal, 542, 1100,
  \dodoi{10.1086/317024}

\bibitem[{{Soler} {et~al.}(2019){Soler}, {Terradas}, {Oliver}, \&
  {Ballester}}]{soler.et19}
{Soler}, R., {Terradas}, J., {Oliver}, R., \& {Ballester}, J.~L. 2019, \apj,
  871, 3, \dodoi{10.3847/1538-4357/aaf64c}

\bibitem[{{Sow Mondal} {et~al.}(2022){Sow Mondal}, {Klimchuk}, \&
  {Sarkar}}]{sowmondal.et22}
{Sow Mondal}, S., {Klimchuk}, J.~A., \& {Sarkar}, A. 2022, \apj, 937, 71,
  \dodoi{10.3847/1538-4357/ac879b}

\bibitem[{Sterling(2000)}]{sterling00}
Sterling, A.~C. 2000, Solar Physics, 196, 79, \dodoi{10.1023/A:1005213923962}

\bibitem[{{Sterling}(2018)}]{sterling18}
{Sterling}, A.~C. 2018, in Journal of Physics Conference Series, Vol. 1100,
  Journal of Physics Conference Series, 012024,
  \dodoi{10.1088/1742-6596/1100/1/012024}

\bibitem[{{Sterling}(2021)}]{sterling21}
{Sterling}, A.~C. 2021, in Solar Physics and Solar Wind, ed. N.~E. {Raouafi} \&
  A.~{Vourlidas}, Vol.~1, 221, \dodoi{10.1002/9781119815600.ch6}

\bibitem[{Sterling {et~al.}(2010)Sterling, Harra, \& Moore}]{sterling.et10a}
Sterling, A.~C., Harra, L.~K., \& Moore, R.~L. 2010, Astrophysical Journal,
  722, 1644, \dodoi{10.1088/0004-637X/722/2/1644}

\bibitem[{Sterling \& Moore(2016)}]{sterling.et16a}
Sterling, A.~C., \& Moore, R.~L. 2016, Astrophysical Journal, 828, L9,
  \dodoi{10.3847/2041-8205/828/1/L9}

\bibitem[{{Sterling} \& {Moore}(2020)}]{sterling.et20a}
{Sterling}, A.~C., \& {Moore}, R.~L. 2020, \apjl, 896, L18,
  \dodoi{10.3847/2041-8213/ab96be}

\bibitem[{Sterling {et~al.}(2015)Sterling, Moore, Falconer, \&
  Adams}]{sterling.et15}
Sterling, A.~C., Moore, R.~L., Falconer, D.~A., \& Adams, M. 2015, Nature, 523,
  437, \dodoi{10.1038/nature14556}

\bibitem[{Sterling {et~al.}(2018)Sterling, Moore, \& Panesar}]{sterling.et18}
Sterling, A.~C., Moore, R.~L., \& Panesar, N.~K. 2018, Astrophysical Journal,
  864, 68, \dodoi{10.3847/1538-4357/aad550}

\bibitem[{{Sterling} {et~al.}(2022{\natexlab{a}}){Sterling}, {Moore}, \&
  {Panesar}}]{sterling.et22}
{Sterling}, A.~C., {Moore}, R.~L., \& {Panesar}, N.~K. 2022{\natexlab{a}},
  \apj, 927, 127, \dodoi{10.3847/1538-4357/ac473f}

\bibitem[{{Sterling} {et~al.}(2023{\natexlab{a}}){Sterling}, {Moore}, \&
  {Panesar}}]{sterling.et23b}
---. 2023{\natexlab{a}}, arXiv e-prints, arXiv:2310.14109,
  \dodoi{10.48550/arXiv.2310.14109}

\bibitem[{{Sterling} {et~al.}(2023{\natexlab{b}}){Sterling}, {Moore},
  {Panesar}, {Samanta}, {Tiwari}, \& {Savage}}]{sterling.et23a}
{Sterling}, A.~C., {Moore}, R.~L., {Panesar}, N.~K., {et~al.}
  2023{\natexlab{b}}, Frontiers in Astronomy and Space Sciences, 10, 49,
  \dodoi{10.3389/fspas.2023.1117870}

\bibitem[{{Sterling} {et~al.}(2020){Sterling}, {Moore}, {Samanta}, \&
  {Yurchyshyn}}]{sterling.et20b}
{Sterling}, A.~C., {Moore}, R.~L., {Samanta}, T., \& {Yurchyshyn}, V. 2020,
  \apjl, 893, L45, \dodoi{10.3847/2041-8213/ab86a5}

\bibitem[{{Sterling} {et~al.}(2022{\natexlab{b}}){Sterling}, {Schwanitz},
  {Harra}, {Raouafi}, {Panesar}, \& {Moore}}]{sterling.et22a}
{Sterling}, A.~C., {Schwanitz}, C., {Harra}, L.~K., {et~al.}
  2022{\natexlab{b}}, \apj, 940, 85, \dodoi{10.3847/1538-4357/ac9960}

\bibitem[{{Uritsky} {et~al.}(2023){Uritsky}, {Karpen}, {Raouafi}, {Kumar},
  {DeVore}, \& {Deforest}}]{uritsky.et23}
{Uritsky}, V.~M., {Karpen}, J.~T., {Raouafi}, N.~E., {et~al.} 2023, \apjl, 955,
  L38, \dodoi{10.3847/2041-8213/acf85c}

\bibitem[{{Uritsky} {et~al.}(2017){Uritsky}, {Roberts}, {DeVore}, \&
  {Karpen}}]{uritsky.et17}
{Uritsky}, V.~M., {Roberts}, M.~A., {DeVore}, C.~R., \& {Karpen}, J.~T. 2017,
  \apj, 837, 123, \dodoi{10.3847/1538-4357/aa5cb9}

\bibitem[{van Ballegooijen \& Martens(1989)}]{vanball.et89}
van Ballegooijen, A.~A., \& Martens, P. C.~H. 1989, Astrophysical Journal, 343,
  971, \dodoi{10.1086/167766}

\bibitem[{{Veronig} {et~al.}(2002){Veronig}, {Temmer}, {Hanslmeier}, {Otruba},
  \& {Messerotti}}]{veronig.et02}
{Veronig}, A., {Temmer}, M., {Hanslmeier}, A., {Otruba}, W., \& {Messerotti},
  M. 2002, \aap, 382, 1070, \dodoi{10.1051/0004-6361:20011694}

\bibitem[{Wang {et~al.}(1998)Wang, Sheeley, Jr., Socker, G., Howard, Brueckner,
  Michels, Moses, Cyr, C., Llebaria, \& Delaboudinière}]{wang.et98}
Wang, Y.-M., Sheeley, N.~R., Jr., {et~al.} 1998, Astrophysical Journal, 508,
  899, \dodoi{10.1086/306450}

\bibitem[{Wyper {et~al.}(2017)Wyper, Antiochos, \& DeVore}]{wyper.et17}
Wyper, P.~F., Antiochos, S.~K., \& DeVore, C.~R. 2017, Nature, 544, 452,
  \dodoi{10.1038/nature22050}

\bibitem[{{Yashiro} {et~al.}(2004){Yashiro}, {Gopalswamy}, {Michalek}, {St.
  Cyr}, {Plunkett}, {Rich}, \& {Howard}}]{yashiro.et04}
{Yashiro}, S., {Gopalswamy}, N., {Michalek}, G., {et~al.} 2004, Journal of
  Geophysical Research (Space Physics), 109, A07105,
  \dodoi{10.1029/2003JA010282}

\bibitem[{Yokoyama \& Shibata(1995)}]{yokoyama.et95}
Yokoyama, T., \& Shibata, K. 1995, Nature, 375, 42, \dodoi{10.1038/375042a0}

\bibitem[{{Young}(2015)}]{young15}
{Young}, P.~R. 2015, \apj, 801, 124, \dodoi{10.1088/0004-637X/801/2/124}

\bibitem[{Young \& Muglach(2014{\natexlab{a}})}]{young.et14a}
Young, P.~R., \& Muglach, K. 2014{\natexlab{a}}, Solar Physics, 289, 3313,
  \dodoi{10.1007/s11207-014-0484-z}

\bibitem[{Young \& Muglach(2014{\natexlab{b}})}]{young.et14b}
---. 2014{\natexlab{b}}, Publications of the Astronomical Society of Japan, 66,
  12, \dodoi{10.1093/pasj/psu088}

\bibitem[{{Zhelyazkov} \& {Chandra}(2018)}]{zhelyazkov.et18}
{Zhelyazkov}, I., \& {Chandra}, R. 2018, \mnras, 478, 5505,
  \dodoi{10.1093/mnras/sty1354}

\bibitem[{{Zhukov} {et~al.}(2021){Zhukov}, {Mierla}, {Auch{\`e}re}, {Gissot},
  {Rodriguez}, {Soubri{\'e}}, {Thompson}, {Inhester}, {Nicula}, {Antolin},
  {Parenti}, {Buchlin}, {Barczynski}, {Verbeeck}, {Kraaikamp}, {Smith},
  {Stegen}, {Dolla}, {Harra}, {Long}, {Sch{\"u}hle}, {Podladchikova}, {Aznar
  Cuadrado}, {Teriaca}, {Haberreiter}, {Katsiyannis}, {Rochus}, {Halain},
  {Jacques}, \& {Berghmans}}]{zhukov.et21}
{Zhukov}, A.~N., {Mierla}, M., {Auch{\`e}re}, F., {et~al.} 2021, \aap, 656,
  A35, \dodoi{10.1051/0004-6361/202141010}

\end{thebibliography}

\clearpage


\begin{figure}
\centering
\includegraphics[width=\textwidth,angle=0]{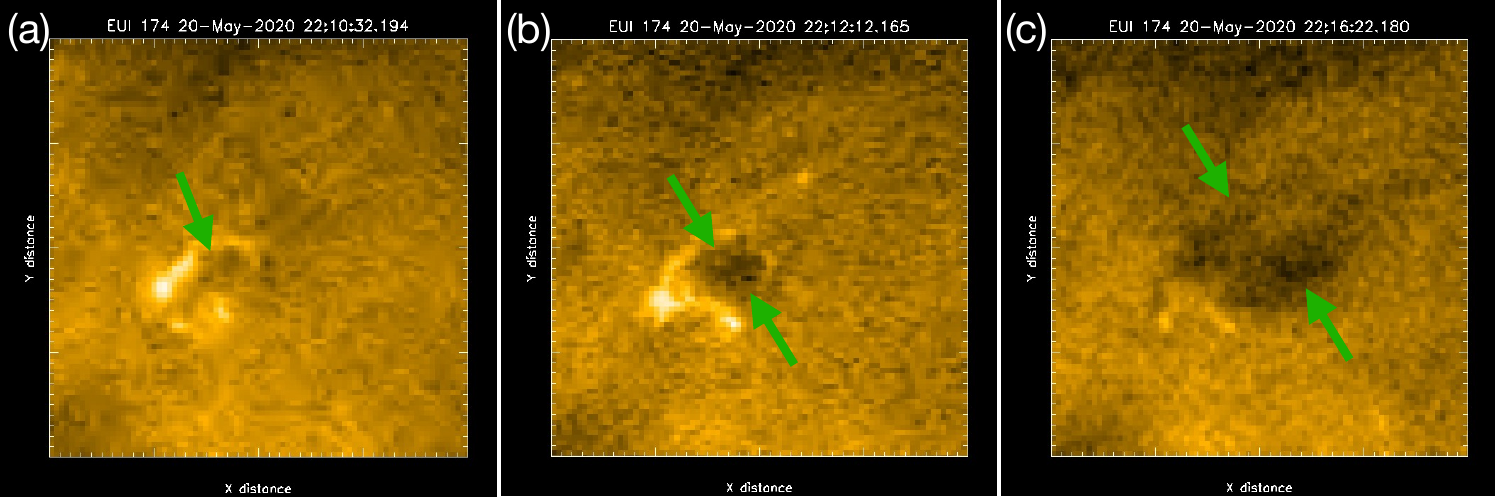}
\caption{\so\ EUV HRI$_{\rm EUV}$~174\,\AA\ images of a ``campfire," as reported by \citet{panesar.et21}.  Arrows show an absorption
feature in the process of erupting. The eruption is essentially identical to the minifilament eruptions that make coronal jets. This is likely 
cool-temperature
material, due to its prominent visibility in cool \sdo/AIA channels \citep[][Fig.\,6]{panesar.et21} In this case, the cool 
material is expelled from the campfire location through (a), (b), and (c).  For both axes, each minor tick represents a distance of about 725\,km.}
\label{eui_nkp_campfire13_crop}  
\end{figure}
\clearpage

\begin{figure}
\centering
\includegraphics[width=12cm,angle=0]{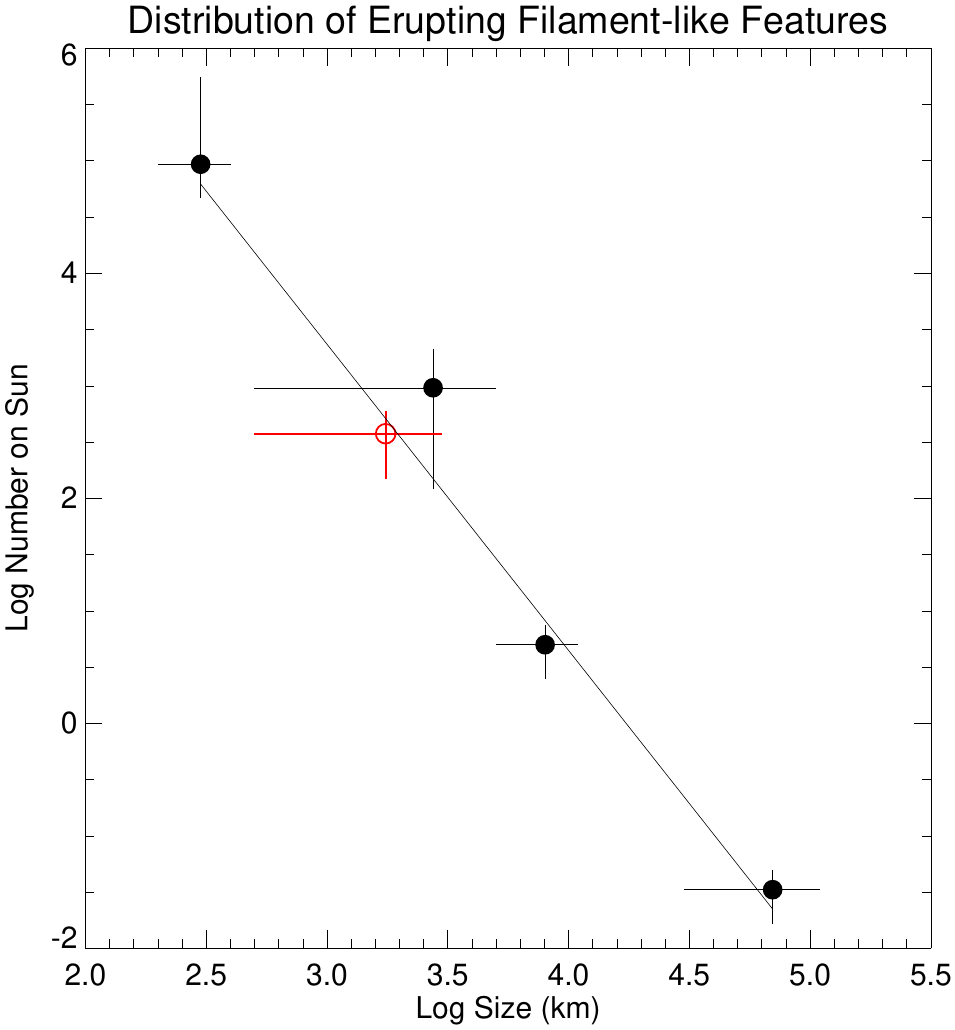}
\caption{Augmentation of a plot in  \citet{sterling.et16a},, showing the number-against-size distribution of erupting filament-like 
features on the Sun. The horizontal axis shows the size (horizontal extent) of the erupting feature, and the vertical axis shows the number 
of events that are erupting on the Sun at any given instant in time.  Among the black-dot values, the point furtherest to the right 
represents solar filaments 
erupting to make typical solar flares and CMEs; the second-from-right represents the minifilaments erupting to make coronal jets; and the 
left-most black dot represents the putative microfilaments erupting to make spicules or spicule-sized features.  These
three points were shown in \citet{sterling.et16a}, and the solid line is a best fit to those three points.  The fourth black dot, which is the
second from the left, is a new addition that represents jetlets on the Sun, based on rate numbers from \citet{raouafi.et23}, and 
the widths of jetlet spires from \citet{panesar.et18b,panesar.et19} and \citet{kumar.et22}.  The red open-circle point represents the number
of outflow events of about jetlet size occurring at a given time over the entire Sun, based on our estimates from the events
 observed by \citet{uritsky.et23} (see text).  Rather than error bars, for each point 
the horizontal and vertical bars represent the approximate range of reported size and rate values, and the solid dot (or red circle) marks the
middle of those ranges.}
\label{fourpt_dist_zu}  
\end{figure}
\clearpage

\begin{figure}
\centering
\includegraphics[width=\textwidth,angle=0]{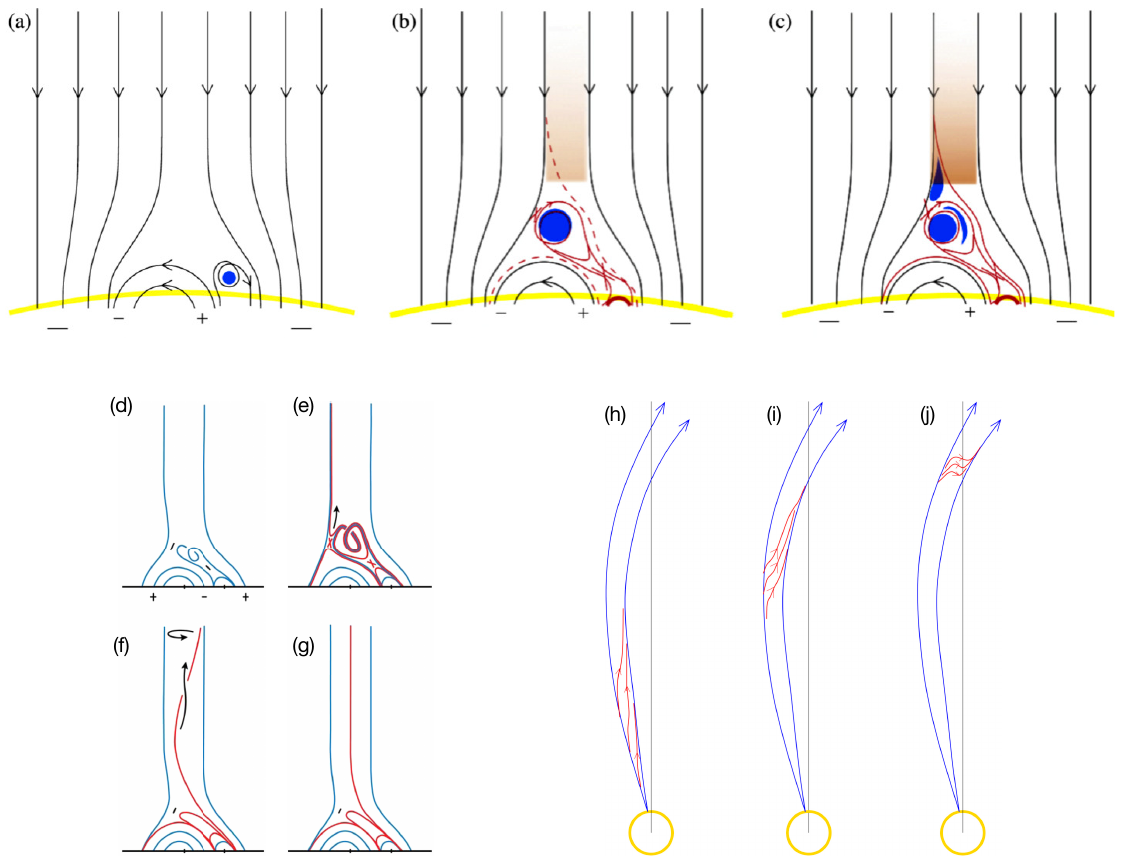}
\caption{Summary of evolution of jetting events into switchbacks and solar wind.  Panels (a)---(c) show the onset
of jets in the minifilament eruption model.  Black lines represent magnetic fields, with the background field a single polarity
(negative) and open into the heliosphere.  The blue circle represents a minifilament, and the looped black line around it indicates
that it resides inside of a strongly sheared field or twisted magnetic flux rope. In (b) the minifilament
is erupting, and its magnetic field is undergoing external reconnection (upper red `x') and internal reconnection (lower red `x'). Red solid lines 
are heated reconnected field lines from the internal reconnection, dashed red lines are reconnected field lines from the 
external reconnection, the 
bright red arc represents the JBP, and the shaded region represents reconnection-heated material flowing outward and forming
the jet spire.  In (c), the external reconnection has progressed enough for the minifilament material to be also flowing out along the 
spire. This is a version of figures in \citet{sterling.et15} and \citet{sterling.et18}, and further details appear in the text and captions of
those papers. Panels (d)---(g) are version of a figure in \citet{moore.et15}, and shows a continuation of the first three panels whereby
the external reconnection of the erupting minifilament's flux rope proceeds until the entire flux rope has opened up onto the open 
field. The twist of the flux rope escapes out into the corona, moving out as an \al ic twist pulse.  Panels~(h)---(j) show a modification of
a figure in \citet{sterling.et20a}. These show a continuation of 
the previous panels, with the magnetic twist wave forming a packet that propagates out into the heliosphere, steepening to become a 
switchback.}
\label{s_m20_matome_zu}  
\end{figure}
\clearpage



\end{document}